\def\mathrm{\rm}
\documentstyle[epsf]{l-aa}
\def\infig#1#2#3{\epsfxsize=#3cm \centering{\mbox{\epsfbox{#2}}}\vspace{-0.4cm}}

\begin{document}

   \thesaurus{06         
              (08.02.2;  
               08.02.4;  
               08.03.2;  
               08.06.3;  
               08.09.2;  
               08.09.2;  
               08.09.2;  
               08.09.2)  
             }
\title{Multiplicity among peculiar A stars I. The Ap stars HD~8441 and
HD~137909, and the Am stars HD~43478 and HD~96391\thanks{Based on observations 
made at the Observatoire de Haute Provence (CNRS), France, at the Jungfraujoch 
station of Observatoire de Gen\`{e}ve, Switzerland, and on data from the ESA 
Hipparcos satellite}$^,$\thanks{Tables 6 to 12 are only available in electronic 
form at the CDS via anonymous ftp to cdsarc.u-strasbg.fr (130.79.128.5)
or via http://cdsweb.u-strasbg.fr/Abstract.html}
}
\author{P. North\inst{1}\and J.-M. Carquillat\inst{2}\and N. Ginestet\inst{2}
\and F. Carrier\inst{1} \and S. Udry\inst{3}}

\institute{Institut d'Astronomie de l'Universit\'e de Lausanne, 
              CH-1290 Chavannes-des-bois, Switzerland
\and  Observatoire Midi Pyr\'en\'ees, 14 avenue Edouard Belin, F-31400 Toulouse,
 France
\and Observatoire de Gen\`eve, CH-1290 Sauverny, Switzerland
	}
\offprints{P.~North}
\date{Received 15 October 1997 / Accepted 25 November 1997}
\maketitle

\markboth{P. North et al.: Multiplicity among Ap, Am stars I}
{P. North et al.: Multiplicity among Ap, Am stars I}

\begin{abstract}
We present the first results of a radial-velocity survey of cool Ap and Am 
stars. HD 8441 is not only a double system
with P = 106.357 days, but is a triple one, the third companion
having an orbital period larger than 5000 days. Improved
orbital elements are given for the classical Ap star HD 137909 = $\beta$ CrB by 
combining our radial velocities with published ones. We yield new orbital 
elements of the two Am, SB2 binaries HD 43478 and HD 96391. Good estimates of 
the individual masses of the components of HD 43478 can be given thanks to the 
eclipses of this system, for which an approximate photometric solution is also 
proposed.
\end{abstract}

\section{Introduction}

Multiplicity plays an essential, though as yet poorly understood, role in the 
formation and maybe the evolution of A-type chemically peculiar stars. While
the rate of binaries tends to be rather small among the magnetic Ap stars, with 
an especially strong deficit of SB2 binaries, it is very high among the Am 
stars, which on the contrary, are
very often found in SB2 systems. There is also a conspicuous lack of orbital
periods smaller than 3 days among the magnetic Ap stars, as shown e.g. in the
review by Gerbaldi et al. (1985). The multiplicity of Am stars has been explored
by Abt \& Levy (1985), and their work showed that all Am stars are not 
necessarily found in binaries (contrary to what earlier results suggested),
or at least not in short-period binaries (even though ``short'' means
$P < 1000$ days).
This indicates that duplicity is not an absolute prerequisite for the Am 
peculiarity to appear, so that the slow rotation of these stars, which 
presumably allows radiative diffusion to work in their atmosphere (Michaud 
et al. 1983), is not always due to tidal friction (Abt 1985, Abt \& Levy 1985).

In order to increase the as yet insufficient statistics, a radial velocity 
survey of cool, magnetic Ap stars has been initiated in 1985 using the CORAVEL 
scanner. Some Am stars were also monitored in the course of this programme, for 
ambiguous classification caused them to be considered as Ap stars. Preliminary 
results of this survey have been published by North (1994). Observations of Am 
stars with CORAVEL had been initiated in the early eighties by M. Mayor and W. 
Benz, with the purpose of determining their projected rotational velocities, but
the results were not published. Since this project was not intended to determine
the rate of binaries, only few stars have more than 2 or 3 measurements, and
HD 96391 was one of them.

In 1992, two of us (JMC and NG) began a radial-velocity survey of all northern 
Am stars whose metallic type is cooler than or equal to F2, with the purpose of 
improving our knowledge of their multiplicity. The limit imposed on the spectral
type deduced from the metallic lines was chosen because of CORAVEL's optimum
efficiency for cool stars. The stars HD 43478 and 96391 were included in this
survey and were therefore measured independently by the Geneva-Lausanne 
observers as well as by the Toulouse observers. The latter measured also a few 
Ap stars, and we present here the results based on the merged data of these
common stars.

The observations are briefly described in Section 2 while the results are
presented and discussed in Sections 3 and 4 for the Ap and Am stars
respectively. Extensive use has been made of Renson's (1991) catalogue in this
work.

\section{Observations}

The radial velocity observations were done at Observatoire de Haute-Provence 
with the CORAVEL scanner attached to the 1-meter Swiss telescope. Although this 
instrument is optimized for late-type stars, it can still yield very good 
results on slowly rotating F stars, and even on A stars if their metallic 
lines are enhanced, as is the case of Ap and Am stars.
Some Ap and Am stars have been measured as early as 1980, but more systematic 
surveys began in 1985 and in 1992 for the Ap and Am stars respectively.

The photometric observations were made in the Geneva system at
different observing sites
(Observatoire de Haute-Provence, Gornergrat and Jungfraujoch),
but most observations of HD 43478 were made at Jungfraujoch (Switzerland)
with the 76-cm telescope.

The individual radial velocities of all stars, as well as individual photometric
measurements of HD 43478, are listed in Tables 6 to 12.

\section{Results on Ap stars}

\subsection{HD 8441 (= BD +42\degr 293 = Renson 2050)}

This bright A2 Sr star was already known as an SB1 system. Its rotational 
period, known from its photometric variability, is 69.43 days
(Rakosch \& Fiedler 1978).
Renson (1966) found an orbital period of 106.3 days and published the
radial-velocity curve, but not the orbital parameters. A total of 107 
measurements have been made over almost 5000
days (Table~6), which confirms the 106 days period (see Fig.~1). However, the 
residuals are larger than expected from the precision of the measurements, and 
follow a very clear trend (see Fig.~2). The presence of a third component is 
certain, although its period is so long that we could not cover even one cycle. 
The orbital parameters of the primary are given in Table~1. This is the second 
spectroscopic triple system known among Ap stars, after the SiMg star HD 201433
whose periods are much shorter (see the catalogue of Tokovinin 1997).
\begin{figure}[th!]
\infig{8.8}{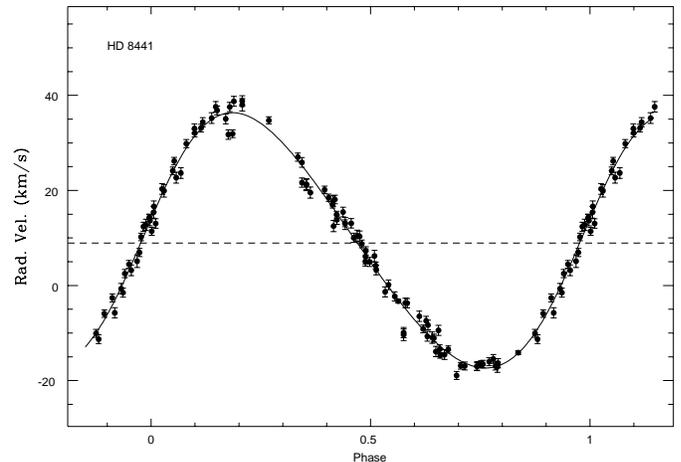}{8.8}
\caption{Radial-velocity curve of HD 8441.
The period is $106.357 \pm 0.009$ days.}
\end{figure}
\begin{figure}[th!]
\infig{8.8}{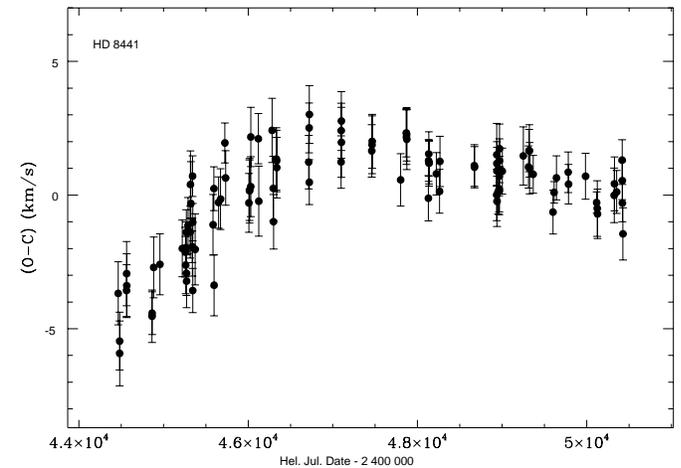}{8.8}
\caption{Radial-velocity residuals vs time for HD 8441.}
\end{figure}
\begin{table*}
\caption{Orbital parameters of the binaries. For each component, the
second line gives the estimated standard deviations of the
parameters}
\begin{center}
\begin{tabular}{|r|r|r|r|r|r|r|r|r|r|r|} \hline
\multicolumn{1}{|c|}{Star name} & \multicolumn{1}{c|}{$P$} &
 \multicolumn{1}{c|}{$T_\circ$ (HJD} & \multicolumn{1}{c|}{$e$} &
 \multicolumn{1}{c|}{$V_\circ$} & \multicolumn{1}{c|}{$\omega_1$} &
 \multicolumn{1}{c|}{$K_{1,2}$} & \multicolumn{1}{c|}{${\cal
 M}_{1,2}\sin^3 i$} & \multicolumn{1}{c|}{$a_{1,2}\sin i$} &
 \multicolumn{1}{c|}{$N$} & \multicolumn{1}{c|}{(O$-$C)} \\ &
 \multicolumn{1}{c|}{(days)} & \multicolumn{1}{c|}{$-2400000)$} & &
 \multicolumn{1}{c|}{($\mathrm km\,s^{-1}$)} &
 \multicolumn{1}{c|}{($^\circ$)} &
 \multicolumn{1}{c|}{($\mathrm km\,s^{-1}$)} & 
 \multicolumn{1}{c|}{$f_{1}$($\cal M$)} &
 \multicolumn{1}{|c|}{$10^{6}$~
 km} & & \multicolumn{1}{c|}{$\mathrm km\,s^{-1}$} \\ \hline HD 8441 
 & 106.357
 & 44952.21 & 0.122 & 8.94 & 279.76 & 26.85 & 0.209 & 38.98 & 107 &
 1.86\\ & 0.009 & 1.32 & 0.010 & 0.19 & 4.75 & 0.29 & 0.007 & 0.42 & & \\ \hline
HD 43478  & 5.464086
 & 47000.1758 & 0 & -6.63 & 0 & 86.48 & 1.777 & 6.498 & 56 &
 1.95\\ & 0.000011 & 0.0037 & fixed & 0.18 & fixed & 0.34 & 0.015 & 0.025 & & \\
 & & & & & & & & & & \\ & & & & & 180 & 95.03 & 1.617 & 7.14 & 57
 & \\ & & & & & fixed & 0.35 & 0.014 & 0.03 & & \\ \hline HD 96391 
 & 4.915427
 & 45234.2422 & 0 & -1.67 & 0 & 84.69 & 1.408 & 5.724 & 44 &
 1.70\\ & 0.000008 & 0.0061 & fixed & 0.20 & fixed & 0.37 & 0.015 & 0.025 & & \\ 
 & & & & & & & & & & \\ & & & & & 180 & 90.19 & 1.322 & 6.10 & 36
 & \\ & & & & & fixed & 0.46 & 0.014 & 0.03 & & \\ \hline HD 137909 
 & 3831.50
 & 44421.40 & 0.5430 & -21.48 & 180.56 & 9.45 & 0.1987 & 417.98 & 78 &
 0.45\\ & 7.94 & 7.63 & 0.0069 & 0.06 & 1.05 & 0.09 & 0.0063 & 4.52 & & \\
& & & & & & & & & & \\  Whole sample 
 & 3858.13
 & 25119.6 & 0.5219 & -21.81 & 180.37 & 9.21 & 0.1943 & 416.81 & 316 &
 0.93\\ & 2.96 & 17.1 & 0.0064 & 0.06 & 1.15 & 0.08 & 0.0058 & 4.13 & & \\
& & & & & & & & & & \\ $P$ fixed & 3858.13
 & 40545.39 & 0.5340 & -21.49 & 180.87 & 9.31 & 0.1955 & 417.66 & 78 &
0.47\\ & fixed & 7.45 & 0.0067 & 0.06 & 1.14 & 0.07 & 0.0056 & 3.96& & \\ \hline
\end{tabular}
\end{center}
\end{table*}
\begin{table}
\caption{Visual magnitude and $v\sin i$ of the programme stars.}
\begin{center}
\begin{tabular}{l|lrr} \hline
\multicolumn{1}{c|}{Star name} & \multicolumn{1}{c}{V}
& \multicolumn{1}{c}{$v$ $\sin$$i$ A} &
 \multicolumn{1}{c}{$v$ $\sin$$i$ B} \\ 
& & \multicolumn{1}{c}{($\mathrm km\,s^{-1}$)} &
 \multicolumn{1}{c}{($\mathrm km\,s^{-1}$)} \\ \hline
 HD 8441  & 6.691& $\leq 2.35\pm 0.60$ & \\
 HD 137909& 3.670& $\leq 7.72\pm 0.15$ & \\
 HD 43478 & 7.483& 28.11 $\pm$ 2.81 & 20.61 $\pm$ 2.06  \\
 HD 96391 & 7.08 & 23.54 $\pm$ 2.35 & 18.04 $\pm$ 0.67 \\ \hline
\end{tabular}
\end{center}
\end{table}
The projected rotational velocity estimated from the width of the 
autocorrelation dip (Benz \& Mayor 1984) is given in Table 2,
with the restriction that in principle,
such a quantity can only represent an upper limit to the true $v\sin i$.
Indeed, the magnetic field commonly present among Ap Sr stars broadens the
lines through the Zeeman effect, so that $v\sin i$ will be overestimated
if this effect is neglected. In this particular case, however, the estimated
$v\sin i$ is quite compatible with the 69.43 days rotational period, assuming
a radius $R\sim 3 ~ R_\odot$.

The Hipparcos parallax of this star is $4.91\pm 0.80$ mas
(Perryman et al. 1997); this translates into a distance $d=232$ pc, after having
applied a Lutz-Kelker correction (Lutz \& Kelker 1973) $\Delta M=-0.28$ which
takes into account the exponential decrease of stellar density in the 
direction perpendicular to the galactic plane. On the other hand, the visual 
absorption estimated from Geneva photometry is $A_v = 0.13$. Assuming a 
contribution of about 0.23 magnitudes of the companions to the visual magnitude 
of the system, the apparent magnitude of the primary alone is $V=6.92$, and 
finally we obtain an absolute magnitude $M_V=-0.03\pm 0.42$ for this component.
Adopting $T_{\mathrm eff}=9200$~K (Adelman et al. 1995) and interpolating in
the evolutionary tracks of Schaller et al. (1992) for a solar metallicity
$Z=0.018$\footnote{This choice of a solar metallicity may appear surprising at 
first sight for Ap and Am stars, whose atmospheric composition is precisely
far from solar. However, it is generally admitted that chemical peculiarities
are confined to the superficial layers of the star (through radiative diffusion)
so that the metal content integrated over the whole stellar mass is the same
as for normal stars, and the deep internal structure remains roughly normal.
But it is true that standard evolutionary tracks should be considered as a first
approximation only for Ap and Am stars; they are also the only ones available
as yet, though Michaud \& Richer (1997) have recently computed fully consistent
evolutionary tracks for F stars, including the effect of radiative diffusion
on the internal stellar structure.}
(and for a moderate overshooting distance $d_{over}/H_p = 0.2$),
one obtains ${\cal M}_1=2.76\pm 0.18 {\cal M}_\odot$, $\log g = 3.71\pm 0.12$ 
($g$ in $cgs$ units) and $R = 3.86\pm 0.66 \,R_\odot$.
Although the uncertainties are fairly 
large, the primary is evolved and on the verge of leaving the Main Sequence; it 
is satisfying that our small $\log g$ value agrees with the spectroscopic
estimate of Adelman et al. (1995) who gave $\log g = 3.35 - 3.8$.

\begin{figure}[th!]
\infig{8.8}{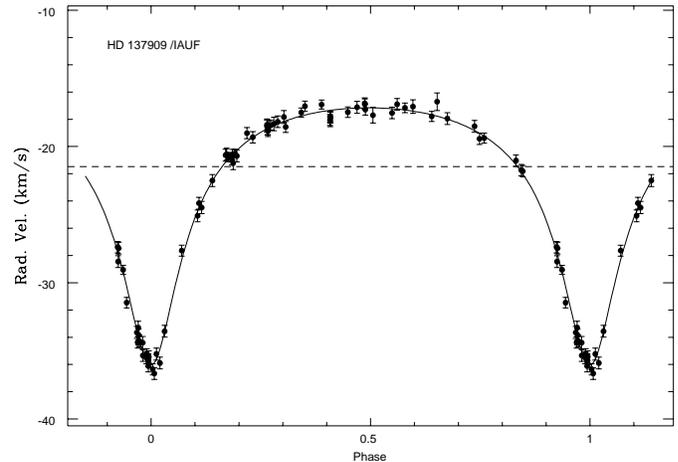}{8.8}
\caption{Radial-velocity curve of HD 137909 given by CORAVEL only. The period is
$3831.50 \pm 7.94$ days.}
\end{figure}
\subsection{$\beta$ CrB (= HD 137909 = BD +29\degr 2670 = Renson 39200)}
This is a well-known, prototype cool Ap star classified A9 SrEuCr.
Its rotational period, known from photometric, spectroscopic and magnetic
variations is 18.4868 days (Leroy 1995). It is known 
as a binary, by both spectroscopy and speckle interferometry. A radial-velocity 
curve was published by Kamper et al. (1990) together with an astrometric orbit 
based on speckle observations. These authors suspected that a third body might 
be present, on the basis of radial velocities taken at Lick Observatory between 
1930 and 1943.
The system was monitored with CORAVEL
for a little more than one cycle, which is very long, 
and 78 measurements have been obtained (see Table~7). The $V_r$ curve is shown 
in Fig.~3 and the spectroscopic orbit is given in Table~1.
Thanks to the precision and the homogeneity of the data, our $V_r$ curve is more
precise than that based on the data taken at David Dunlap Observatory by
Kamper et al. 
Combining our measurements with those published by Kamper et al. 
(1990), by Oetken \& Orwert (1984) and by Neubauer (1944), we can refine
the period to $P = 3858.13$ days, but the accuracy of the orbital elements is
not improved, due to the scatter of the residuals, which is more than twice 
larger than for CORAVEL observations alone. In order to fit Neubauer's data
to the others, we had to subtract a constant value (2~km\,s$^{-1}$) to them,
which was also done by Kamper et al. (1990). The resulting radial-velocity curve
is shown in Fig.~4 and the corresponding orbital elements are given in Table~1.
The residuals are shown in Figure 5. A fit of the CORAVEL radial velocities
alone has also been done keeping the orbital period fixed to the above, refined
value and its results are displayed in the last line of Table 1. The scatter
of the $O-C$ residuals is hardly increased and only the eccentricity and the
amplitude change by more than one sigma with respect to the fit where $P$ was
adjusted; the change is probably due to the rather inhomogeneous phase coverage
of the observations.

\begin{figure}[th!]
\infig{8.8}{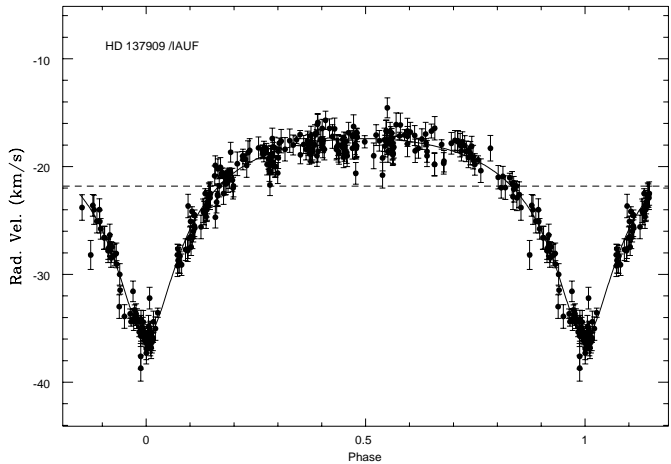}{8.8}
\caption{Radial-velocity curve of HD 137909 including data published by Kamper 
et al. (1990), by Neubauer (1944) and by Oetken \& Orwert (1984).
A correction of $-2 {\rm\, km\,s^{-1}}$ has been added to the $V_r$ values of 
Neubauer. The period is $3858.13 \pm 2.96$ days. }
\end{figure}

\begin{figure}
\infig{8.8}{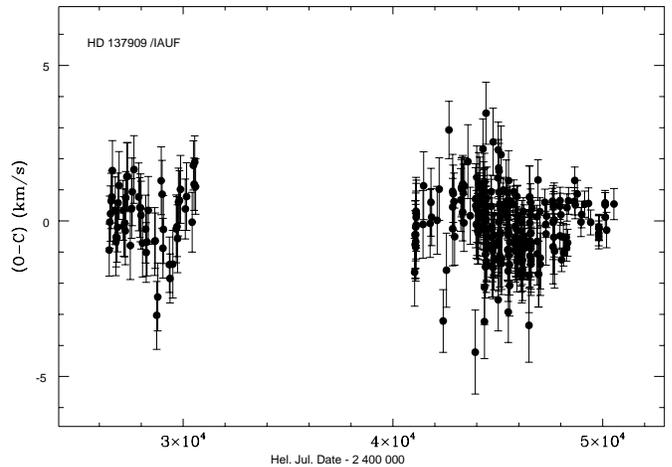}{8.8}
\caption{Radial-velocity residuals vs time for HD 137909 for the whole sample.}
\end{figure}

Thanks to the Hipparcos satellite, $\beta$ CrB has now a very precise parallax 
$\pi = 28.60 \pm 0.69$ mas which allows to compute the linear semi-major axis of
the relative orbit from the angular semi-major axis obtained by speckle 
interferometry ($203.2 \pm 1.4$ mas). Since the inclination angle
$i = 111.11\degr \pm 0.46\degr$ of the orbit is known from speckle 
interferometry and the quantity $a_1 sin$~$i$ is known from CORAVEL 
measurements, the semi-major axis of the absolute orbit of the companion can be 
computed:
\begin{equation}
a_2=a-a_1=4.114 \pm 0.031~U.A. = 615.2 \times 10^6 km
\end{equation}
as well as the mass ratio :
\begin{equation}
{{\cal M}_2\over{\cal M}_1}={a_1\over a_2}={0.727 \pm 0.033}
\end{equation}
Finally, one can obtain ${\cal M}_{2} = 1.356 \pm 0.073 ~{\cal M}_\odot$ from 
the mass function, as well as ${\cal M}_{1} = 1.87 \pm 0.13 ~{\cal M}_\odot$.
Oetken \& Orwert (1984) had found
${\cal M}_{1} = 1.82$ and ${\cal M}_{2} = 1.35$ using the same method but a 
pre-Hipparcos parallax of 31 mas. Our results, altough close to theirs, is more 
reliable. The radius of the primary, estimated from the Hipparcos parallax
and from $T_{\rm eff}= 7750$ K (Faraggiana \& Gerbaldi 1993), is
$R = 3.03\pm 0.25 \,R_\odot$, which implies an equatorial velocity
$v_{eq}= 8.3\pm 0.7$ km\,s$^{-1}$. If the rotational equator of the star
coincides with the orbital plane, then the projected rotational velocity
is $v\sin i = 7.7\pm 0.7$ km\,s$^{-1}$, in excellent agreement with the value
(which may be overestimated, however) listed in Table 2. In these estimates,
we assumed a negligible interstellar absorption and adopted the difference
$\Delta V = 1.7$ mag between the components of this speckle binary
(Tokovinin 1985), so that the apparent visual magnitude of the primary
component alone is 3.876 instead of 3.670 for the whole system (Rufener 1988).

The HR diagram is shown on Figure 6. Strangely enough, the agreement between
the observed location of $\beta$ CrB and the evolutionary track at the observed
dynamical mass is very poor: both the primary and the secondary (if we
rely on $\Delta V = 1.7$) appear overluminous compared to the
evolutionary tracks drawn for their mass. Considered alone, the primary might 
well be at the very tip of the blue hook at the core-hydrogen exhaustion phase,
which would reconcile within one $\sigma$ its observed and theoretical locations
in the HR diagram. Its logarithmic age might then be 9.05 dex instead
of 8.9. However, the secondary (indicated in Figure 6 as a dot arbitrarily
placed along the abcissa on the isochrone $\log t=8.9$) seems overluminous as 
well, making the puzzle more complicated but also more interesting, and 
certainly well worth further investigations. Unfortunately, it is not possible 
to test completely the position of the secondary because of its unknown
colours\footnote{Bonneau \& Foy (1980) give $\Delta m = 1.5$ at 6500~\AA, but
the uncertainty seems too large for the inferred $V-m_{6500}$ index to be really
useful}.
Such an information would be most interesting to test the idea of Hack et al.
(1997) that the companion might be a $\lambda$ Boo star with
$T_{\rm eff}\sim 8200$~K, although such a hypothesis appears difficult to
maintain in view of Figure 6.
\begin{figure}[th!]
\infig{8.8}{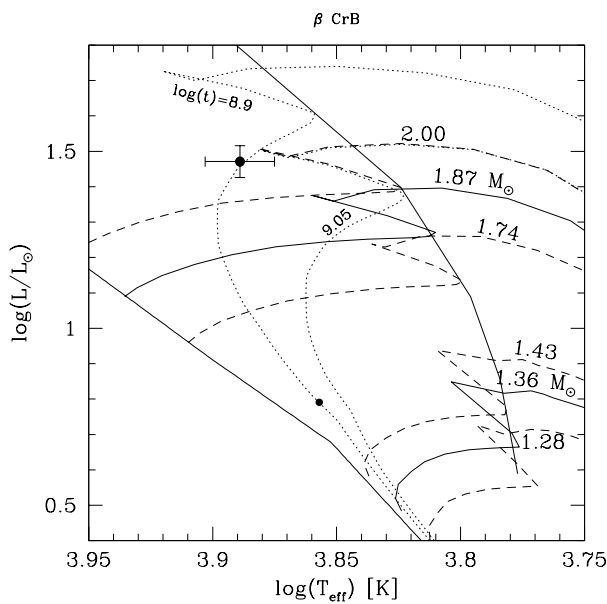}{8.8}
\caption{HR diagram of $\beta$ CrB. The ZAMS, TAMS and evolutionary tracks 
interpolated for the observed masses are shown as solid lines, while those 
interpolated for the masses $\pm 1\ \sigma$ appear as broken lines. The dotted 
lines indicates the isochrones at $\log t = 8.9$ and 9.05 ($t$ in years).}
\end{figure}

The semi-major axis of the orbit of the binary's photocenter is given in the
Hipparcos and Tycho Catalogues (Perryman et al. 1997). This allows an 
independant test of the magnitude difference $\Delta V$: assuming the 
photocenter to be defined by $E_1 x_1 = E_2 x_2$, where $E_1$ and $E_2$ are the
respective brightnesses of the components in the $H_p$ passband and $x_1$, $x_2$
the distances of the components to the photocenter such that 
$x_1+x_2=a_1+a_2=a$, one obtains $a_{\rm o}= 1.77\pm 0.07$~au, on the basis
of Tokovinin's $\Delta V = 1.7$ (one has $a_{\rm o}= a_1-x_1$). This is in
rough agreement (within three sigmas) with $a_{\rm o}= 1.97\pm 0.05$~au given
in the Hipparcos catalogue. Our estimate assumes a companion with
$T_{\rm eff} = 7200$~K and takes into account the colour equation
between $Hp$ and $V$ (Vol. 2, p. 59 of Perryman et al. 1997) which leads to
$\Delta Hp = 1.71$. Increasing $\Delta V$ by about 0.2 magnitudes
would bring perfect agreement. Unfortunately, Tokovinin (1985) do not give any
error estimate on $\Delta V$.

\section{Results on Am stars}

\subsection{HD 43478 (= BD +32\degr 1246 = Renson 11540)}

This star was classified A3-F2-F5 by Osawa (1965) and Ap~Si~Sr by Bertaud \&
Floquet (1974). As kindly pointed by
Renson (1994, personal communication to PN), Babcock (1958) had already
found it double-lined, but did not give any period.
Interestingly, Babcock listed this star as probably magnetic, and mentioned
the profile of the K line as peculiar; as Babel (1994) showed, cool magnetic
Ap stars have a peculiar profile of the Ca\,{\sc ii} K line which betrays a
stratification of calcium in the star's atmosphere. Perhaps this star should
be indeed classified Ap after all. We secured 56 points (see Table~9, 10 and
Fig.~7) and obtained the orbital elements listed in Table~1.

\begin{figure}[th!]
\infig{8.8}{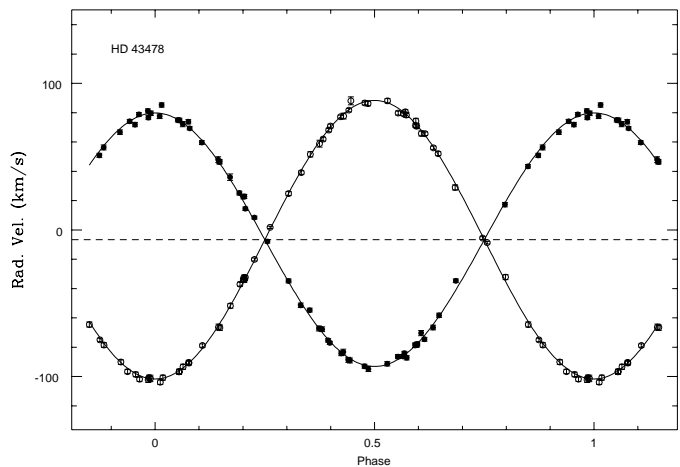}{8.8}
\caption{Radial-velocity curve of HD 43478.
The period is $5.464086 \pm 0.000011$ days. Notice that the zero phase
corresponds here to the quadrature (epoch given in Table 1) but not to the
primary eclipse (Equation 3) which would fall here at phase 0.75, when the
{\it more massive} component passes {\it in front} of the less massive one.}
\end{figure}
\begin{figure*}
\infig{8.8}{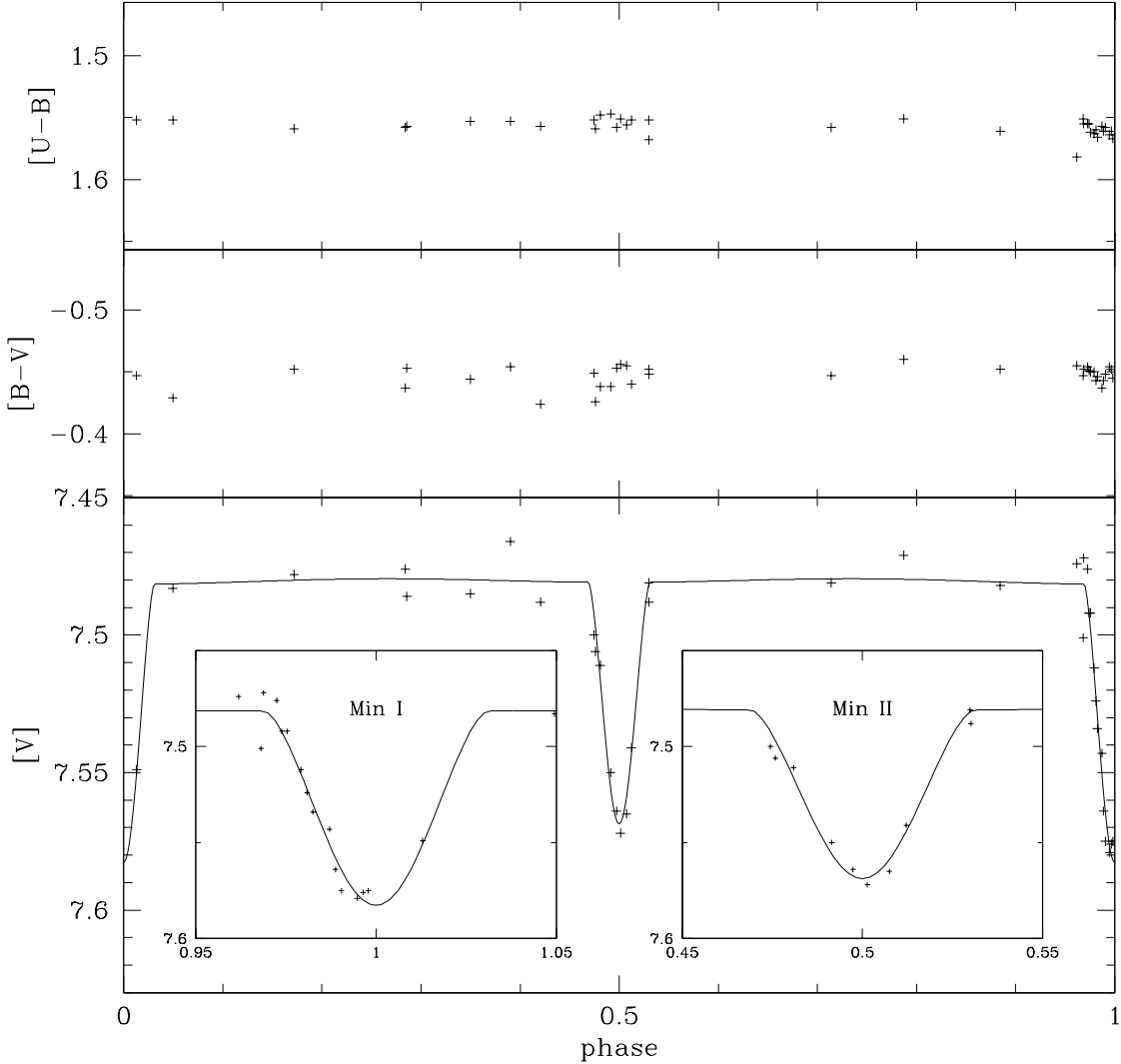}{16}
\caption{Lightcurves of HD 43478 for the $[V]$
magnitude and the $[U-B]$ and $[B-V]$ colour indices of Geneva photometry,
plotted according to the ephemeris given in Equation 3. 
Notice the lack of colour change during eclipses. The best fit to the $[V]$ 
curve is shown (see Table 4).}
\end{figure*}

The system is especially interesting, because we discovered eclipses, which 
allow to determine the orbital inclination (North \& Nicolet 1994). 
Unfortunately, the eclipses are shallow, as shown if Figure~8 where the
lightcurve is plotted according to the ephemeris:
\begin{eqnarray}
HJD(Min I) = 2\,446\,774.790 &+ 5.464086\,E \\
\pm 0.003& \nonumber
\end{eqnarray}
In addition, the number of measurements is small, due to the unfavourable
period (close to 5.5 days) and we had a relatively small number of good nights
at the Jungfraujoch station on the critical dates.
The descending branch of the primary minimum was observed during a mediocre,
partial night where only five standards could be measured; nevertheless,
the scatter around the fitted lightcurve is fairly good.
The two minima have about the same depth and are separated by exactly 0.5 in
phase, confirming the circularity of the orbit. On the other hand, the $[U-B]$ 
and $[B-V]$ curves remain flat during both eclipses, showing that both
components have similar effective
temperatures. In spite of the small number of points, we analized the lightcurve
with the EBOP16 programme (Etzel 1980, 1991). The Geneva and $uvby\beta$ 
photometric indices and parameters give very consistent $T_{\mathrm eff}$ and 
$\log g$ values through the calibration of K\"unzli et al. (1997) for the Geneva
system, and through the calibration of Moon \& Dworetsky (1985) for the 
$uvby\beta$ system (Table~3). For the $uvby\beta$ system, we applied the 
correction recommended by Dworetsky \& Moon (1986) to the $\log g$ value for the
Am stars.

\begin{table*}
\caption{Physical parameters of HD 43478 according to its colours in the
$uvby\beta$ and Geneva photometric systems. Both components are assumed to be 
identical. Note that $E(B2-V1)=1.146 E(b-y)$. The errors quoted for the
physical parameters determined with Geneva photometry are propagated from
typical errors on the colour indices but do not include possible systematic
errors related with the calibration itself. The reddening $E(B2-V1)=0.089$ 
corresponds to $E(B-V)=0.1$ suggested by the maps of Lucke (1978) and is
mentioned only to illustrate its effect on the physical parameteres; the
adopted colour excess $E(B2-V1)=0.045$ is obtained from $E(b-y) = 0.039$,
which results from the calibrated $uvby\beta$ colours.}
\begin{center}
\begin{tabular}{c|rrrrr} \hline
\multicolumn{1}{c|}{Photometry} & \multicolumn{1}{c}{$T_{\rm eff} [K]$} &
 \multicolumn{1}{c}{$\log g [cgs]$} & \multicolumn{1}{c}{$[M/H]$} &
 \multicolumn{1}{c}{$E(b-y)$} &
 \multicolumn{1}{c}{$E(B2-V1)$}\\ \hline $uvby\beta$
 & 7026
 & 3.76 & 0.88 & 0.039 & \\ \hline Geneva & 6862 $\pm$ 51 & 3.78 $\pm$ 0.16 &
0.47 $\pm$ 0.06 & 0.039 & 0.045\\ \\
& 7189 $\pm$ 58 & 4.18 $\pm$ 0.08 & 0.55 $\pm$ 0.07 && 0.089\\ \hline
\end{tabular}
\end{center}
\end{table*}

From the values of $T_{\mathrm eff}$ and $\log g$, we interpolated the linear
limb-darkening coefficient $u$ from the tables of Van Hamme (1993). With the
available photometric data, it is impossible to fit simultaneously all the
interesting parameters, namely the central surface brightness of the secondary
$J_s$, the radius $r_p$ of the primary, the ratio $k$ of the radii and the
orbital inclination $i$. This is a well-known difficulty for all systems
(even well detached ones) where both components are nearly identical, even when
the eclipses are deep. Another type of data has to be used to constrain the
ratio of radii, because the latter may be changed from e.g. 0.6 to 1.4,
without any change in the $rms$ scatter of the residuals. We do not have
detailed spectroscopic informations, but the CORAVEL data allow to have a
rough guess of the $k$ ratio in the two following ways:
\begin{enumerate}
\item The width (FWHM) of the autocorrelation dip can be translated in terms
of $v\sin i$ through a proper calibration (Benz \& Mayor 1984).
Assuming there is no other cause
of broadening than in normal stars (i.e. no Zeeman broadening, for instance),
one obtains in this way the projected rotational velocities given in Table~2.
If synchronism has taken place between spin and orbital periods, which appears
highly probable given the rather evolved state of the system (low $\log g$)
and the circular orbit (circularisation time is longer than synchronisation
time according to tidal theories), then $k$ is directly given by the ratio
of the $v\sin i$ values, i.e. 0.73.
\item The equivalent width $W$ of the autocorrelation dip depends on effective
temperature and metallicity of the star, but also on the amount of dilution
of the stellar flux by the companion's flux. Assuming that both stars have the
same effective temperature (as suggested by the flat $[U-B]$ and $[B-V]$ curves)
and the same metallicity (a more adventurous assumption), the ratio $W_2/W_1$
gives directly the luminosity ratio $L_2/L_1$ and is equal to the square
of the ratio of radii $k^2$. One obtains in this way $k=0.79$.
\end{enumerate}
A larger weight has to be granted to the first method, so we adopt here
$k=0.75$, keeping in mind that the uncertainty on this quantity remains
considerable (20\% or so).
The final elements found with the EBOP16 programme are given in
Table~4. They are rather approximate, but the inclination is relatively well
determined and so are the masses too. It is necessary here to comment briefly
on the definition of ``primary'' and ``secondary'' components, because it is
not necessarily the same when radial velocities, respectively lightcurves are
considered. From the radial-velocity standpoint, the primary evidently
corresponds to the smaller amplitude $K$ and to the more massive component.
But, when interpreting the lightcurve, the EBOP code assumes that the deeper (or
primary) eclipse corresponds to the {\it secondary} passing in front of the
primary component. In this particular system, it is interesting to notice
that the primary minimum corresponds to phase 0.75 of the $V_r$ curve, where
the {\it less} massive component lies {\it behind} the primary, not the reverse.
Therefore, the adopted ratio of radii entered into the EBOP code should not be
$k = r_2/r_1 = 0.75$, but $k = r_s/r_p = 1.333$, since we have to identify the
dynamical primary (1) with the photometric secondary (s) and the dynamical
secondary (2) with the photometric primary (p). Interestingly, this implies
a larger surface brightness of the dynamical secondary than of the primary,
i.e. a slightly larger effective temperature, a relatively rare occurence.
The effective temperatures have been computed from an apparent
$T_{\rm eff} = 6944$ K (average of Geneva and $uvby\beta$ estimates) which
is assumed to result from a weighted average of the components' reciprocal 
temperatures:
$\theta_{\rm eff}(apparent)=0.7258=(L_1\theta_1+L_2\theta_2)/(L_1+L_2)$,
and assuming $J_s/J_p = (T_{\rm eff s}/T_{\rm eff p})^4$.
The bolometric luminosity has been computed assuming $M_{\rm bol\odot}=4.75$.
\begin{table}
\caption{parameters of HD~43478 obtained from the $[V]$ magnitude using
the EBOP16 code and assuming the ratio of radii $r_2/r_1 = 0.75$. The indicated
errors are the formal ones only and do not include the large uncertainty
on $k$. Notice that the subscripts $p$ and $s$ refer to the {\it photometric}
primary and secondary respectively (the ``secondary'' being defined as the
foreground star at Min. I) but correspond to the subscripts 1 and 2 (in this
order), which correspond to the more and less massive star respectively.}
\begin{center}
\begin{tabular}{cl} \hline
 \multicolumn{1}{c}{Parameter}
& \multicolumn{1}{c}{value $\pm \sigma$ ($[V]$ band)}\\ \hline
$i \,[^\circ$] & 78.94  $\pm 0.35$ \\
$r_p=R_p/a = R_2/a$& 0.1157 $\pm 0.0033$ \\
$k=r_s/r_p=R_1/R_2=1/0.75$            & 1.333    \\
$r_s=R_s/a = R_1/a$& 0.1542 $\pm 0.0044$ \\
$u_p = u_2$    & 0.500    \\
$u_s = u_1$    & 0.500    \\
$J_s/J_p = J_1/J_2$& 0.914  $\pm 0.049$  \\
$L_p/(L_p+L_s) = L_2/(L_1+L_2)$& 0.38     \\
$L_s/(L_p+L_s) = L_1/(L_1+L_2)$& 0.62     \\
$\sigma_{\mathrm res}$ [mag]&0.0073  \\ \hline
\end{tabular}
\end{center}
\end{table}

\begin{table}
\caption{Physical parameters of the components of HD 43478. The error on
the masses includes a large, 20\% uncertainty on $k$, which translates into
a $\pm 0.27^{\rm o}$ uncertainty on $i$. The same is true of the 
radii, whose uncertainties are mutually anticorrelated since the sum of
radii remains constant within 3\% as $k$ is varied.}
\begin{center}
\begin{tabular}{lll} \hline
 &\multicolumn{1}{c}{Primary}&\multicolumn{1}{c}{Secondary} \\ \hline
${\cal M}/{\cal M_\odot}$& $1.880 \pm 0.018$ & $1.710 \pm 0.017$ \\
$R/R_\odot$            & $3.08 \pm 0.24$ & $2.31 \pm 0.31$ \\
$\log g$ [cgs]         & $3.735 \pm 0.072$ & $3.94 \pm 0.12$ \\
$v\sin i$ [km s$^{-1}$]& $28.1 \pm 2.8$        & $20.6  \pm 2.1  $ \\
$\log T_{\mathrm eff}$ [K]&$3.838 \pm 0.010$& $3.847\pm 0.010$ \\
$\log L/L_\odot$       & $1.28 \pm 0.11$ & $ 1.07 \pm 0.16$ \\
$M_{bol}$              & $1.55 \pm 0.27$   & $2.08 \pm 0.39 $  \\
$B.C.$                   & $-0.10$           & $-0.10$           \\
$M_v$                  & $1.65 \pm 0.27$   & $2.18 \pm 0.39$   \\
$E(B2-V1)$             & \multicolumn{2}{c}{$0.045 \pm 0.015$} \\
$A_V$                  & \multicolumn{2}{c}{$0.17 \pm 0.06$}  \\
Distance [pc]          & \multicolumn{2}{c}{$172 \pm 22$} \\
Age ($\log t$, $t$ in years)& \multicolumn{2}{c}{$9.10 \pm 0.08$} \\ \hline
\end{tabular}
\end{center}
\end{table}

A summary of the physical parameters of the HD 43478 system is given in
Table~5. The bolometric correction has been taken from Schmidt-Kaler (1982).
The distance is shorter than indicated by the Hipparcos satellite,
which gave $\pi = 3.87\pm 0.93$ mas or $d = 258_{-50}^{+82}$ pc; however,
the discrepancy cannot be considered significant since it remains largely
within two sigmas. The distance deduced from the fundamental radii and
photometric effective temperatures is almost twice more accurate than that
given by Hipparcos (the error has been estimated using the usual propagation
formula applied to the distance modulus). It is interesting that the fundamental
$\log g$ value we find for the primary, which is quite reliable, is in excellent
agreement with the value obtained from both Geneva and $uvby\beta$ colour
indices.

The situation of both components in the HR diagram is
shown in Figure 9, together with $Z=0.020$ evolutionary tracks interpolated
in those of Schaller et al. (1992) and isochrones with ages $\log t = 9.1$
and $9.2$. The primary is clearly at the end of its life on the Main Sequence,
and the secondary is probably somewhat evolved too. In view of the shape
of the isochrones, one can easily understand why the secondary is slightly 
hotter than the primary. Clearly, a more complete and precise
lightcurve is needed, especially to give a more accurate, fundamental estimate 
of the radii and to assess thereby the validity of the assumption of 
synchronism.
\begin{figure}[th!]
\infig{8.8}{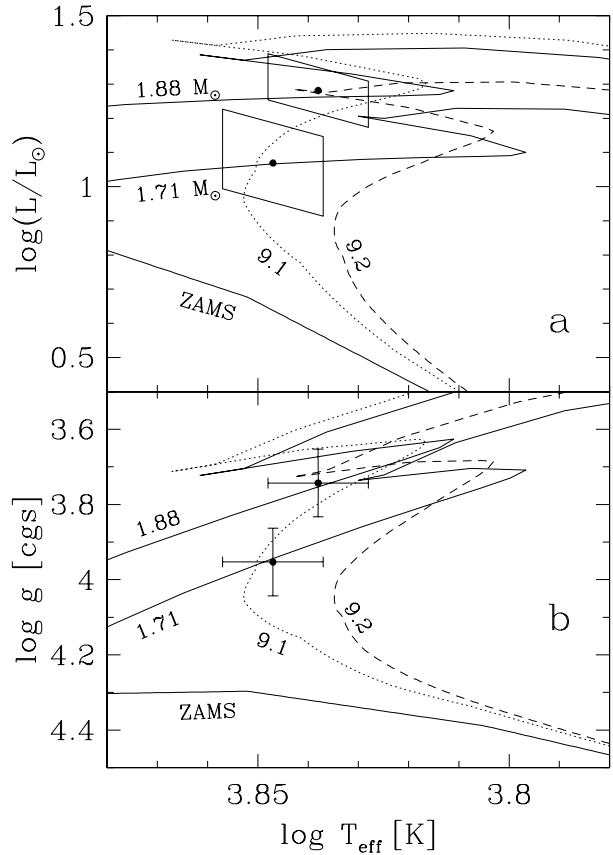}{8.8}
\caption{HR and $\log g$ vs $\log(T_{\mathrm eff})$ diagrams for both
components of HD 43478. The position of the primary is fairly well defined,
while that of the secondary is less reliable. The continuous lines are
the ZAMS and evolutionary tracks interpolated for the measured masses, while
the dotted and broken lines are the isochrones at $\log t = 9.1$ and
$\log t = 9.2$ respectively.}
\end{figure}

\subsection{HD 96391 (= BD +72\degr 515 = Renson 27770)}
\begin{figure}[th!]
\infig{8.8}{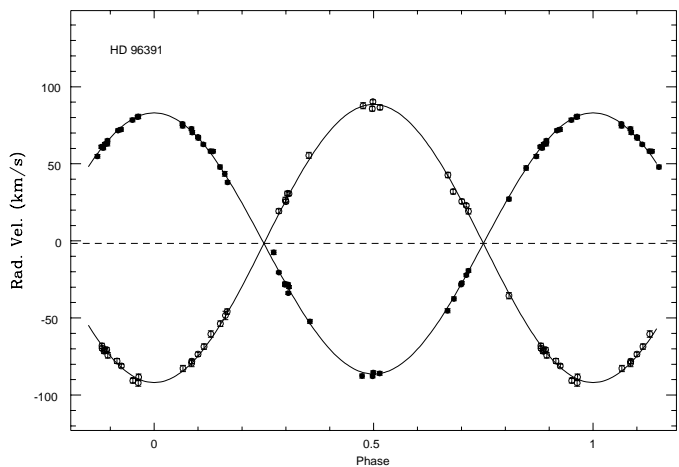}{8.8}
\caption{Radial-velocity curve of HD 96391. The period is
$4.915427 \pm 0.000008$ days. }
\end{figure}
This star was classified A4-F0-F3 by Abt (1984). It is also an SB2 system
with very similar companions. Unfortunately, we do not have Geneva
photometry for that star, but Str\"omgren photometry\footnote{b-y = 0.220,
$m_1$ = 0.225, $c_1$ = 0.656, V = 7.08}
done by Olsen (1983) and retrieved using the General Catalogue of Photometric
Data (Mermilliod et al. 1997) gives $T_{\mathrm eff} = 7020~K$, $\log g = 3.85$
through the calibration of Moon \& Dworetsky (1985), and $\Delta$$m_0 = -0.051$,
$R/R_\odot = 1.74$, $M_V = 2.66$, $M_{bol} = 2.59$ and
$\log (L/L_\odot) = 0.82$ through older calibrations included in Moon's (1985)
code. We have 36 CORAVEL observations of this star (Tables~11 and 12); the
orbital elements are listed in Table~1 and the
$v\sin i$ value of each component is given in Table~2. The $i$
angle remains unknown, since that star is not known
as an eclipsing binary.

On the other hand, the Hipparcos parallax is $\pi = 6.81\pm 0.62$
mas, which implies a distance $d = 152$ pc taking into account the Lutz-Kelker 
correction -0.07. Furthermore,
from the reddening maps of Lucke (1978), a colour excess $E(B-V) = 0.029$
appears reasonable, so we adopt $A_V \sim 0.10$; to correct for the duplicity,
the apparent visual magnitude is increased by 0.75 mag (so the result will
relate to an average component), and one obtains
$M_V=1.82$, $\log(L/L_\odot)=1.198\pm 0.089$, and
by interpolation in theoretical evolutionary tracks, $\log g = 3.84\pm 0.08$
dex, ${\cal M} = 1.819\pm 0.070 {\cal M}_\odot$ and $R = 2.69\pm 0.30 R_\odot$.
The agreement of the $\log g$
value obtained here with that given by the $uvby$ photometry is excellent
(the photometric luminosity is far off, but is obtained through an older
calibration). Once again, this system appears close to the end of its life
on the main sequence.
It is now possible to estimate the orbital inclination $i$
by comparing ${\cal M}\sin^3i \simeq 1.38 {\cal M}_\odot$ with
${\cal M}=1.82 {\cal M}_\odot$ and the result is $i\simeq 66\degr$.
This precludes eclipses, which would need an orbital inclination larger
than $\sim 73\degr$ to occur. The individual masses are about
${\cal M}_1 = 1.85 {\cal M}_\odot$ and ${\cal M}_2 = 1.73 {\cal M}_\odot$.
The average equatorial velocity computed from the radius obtained above
and from the assumption of synchronism is 27.7 km\,s$^{-1}$, which translates
into $v\sin i\simeq 25$ km\,s$^{-1}$. This value may be compared with the
observed $v\sin i$'s of both companions (Table 2), if the spin axes are
perpendicular to the orbital plane: the observed values are smaller than those
predicted by synchronism via the radius estimate, but the $11\%$ error on the
latter is large enough to accomodate both results within two $\sigma$.
The system is very probably synchronised.

\section{Conclusion}

We have shown that the cool Ap star HD 8441 belongs to a triple
system, whose ratio of the long to the short period is larger than 47.

We could improve the knowledge of the orbit of the classical Ap
star $\beta$ CrB thanks to our new, highly
homogeneous radial velocities which could be combined with the
published speckle orbit. Furthermore, the individual masses of the
components could be determined thanks to the Hipparcos parallax.

The masses of both components of the Am star HD 43478 could be
obtained with 1\% accuracy thanks to the eclipsing
nature of the system. The radii are less precisely known, because of
the shallowness of the eclipses and the insufficient photometric data.
Nevertheless, the assumption of synchronism combined with the present data
leads to a very good match of both components with the isochrone $\log t=9.1$,
and the evolved state of at least the primary clearly appears even without
this assumption. It would be worthwhile to make further
photometric observations of this system,
in order to obtain a better estimate of the radii and to tell whether
it is synchronized or not. Spectroscopic observations should also be done
in order to constrain the luminosity ratio and the individual effective
temperatures.

Finally, we obtained a good mass ratio for the Am star HD 96391 and could
also estimate its orbital inclination from an independant mass determination
using the Hipparcos parallax.

All four binaries examined here have at least one component (the primary)
which is significantly evolved and will leave the Main Sequence within
a short time, relatively to its MS lifetime. The rotational period of both
components is very probably synchronised with the orbital period in the
two systems hosting Am stars.

\acknowledgements{This work was supported in part by the Swiss National
Foundation for Scientific Research. Part of the reduction of the data was made
by the late Dr. Antoine Duquennoy. The photometric data were reduced by
Mr. Bernard Pernier, Mr. Christian Richard, Prof. Fr\'edy Rufener and
Prof. Gilbert Burki. We thank Dr. S. Hubrig for drawing our attention to 
Tokovinin's 1985 paper and 1997 catalogue.}

\end{document}